\documentclass{appolb}
\pdfoutput=1
\usepackage{graphicx}
\usepackage{booktabs}
\usepackage{amssymb,amsmath,amsfonts}
\usepackage[utf8]{inputenc}
\usepackage{color}
\usepackage{nicefrac}

\usepackage[square,numbers,sort&compress]{natbib}
\usepackage[breaklinks,hidelinks,colorlinks]{hyperref}
\hypersetup{
    linkcolor={blue},
    citecolor={blue},
    urlcolor={blue}
}

\begin{document}
\title{Modeling chiral criticality%
\thanks{Presented at CPOD 2016 in Wroclaw}%
}
\author{Bengt Friman
\address{GSI Helmholtzzentrum f\"{u}r Schwerionenforschung, D-64291 Darmstadt, Germany}
\and
G\'{a}bor Andr\'{a}s Alm\'{a}si 
\address{GSI Helmholtzzentrum f\"{u}r Schwerionenforschung, D-64291 Darmstadt, Germany}
\and
Krzysztof Redlich
\address{ExtreMe Matter Institute EMMI, GSI, D-64291 Darmstadt, Germany}
\address{University of Wroc\l aw, Institute of Theoretical Physics,   PL 50-204 Wroc\l aw, Poland}
}
\maketitle
\begin{abstract}
We discuss the critical properties of net-baryon-number fluctuations at the   chiral restoration transition
in  matter at non-zero temperature and net-baryon density.
The chiral dynamics of quantum chromodynamics (QCD) is modeled by the Polykov-loop extended  Quark-Meson  Lagrangian,  that includes the coupling of quarks to temporal gauge fields. The Functional Renormalization Group is employed to account for the  criticality at the phase boundary. We focus on the ratios of the net-baryon-number cumulants, $\chi_B^n$, for $1\leq n\leq 4$. The results are confronted with recent experimental data on fluctuations of the net-proton number in nucleus-nucleus collisions.  
\end{abstract}
\PACS{PACS numbers come here}
  
\section{Introduction \label{sec:Introduction}}

Lattice QCD (LQCD) results imply that at  vanishing  and small values of baryon chemical potential,  $\mu_B$,  strongly interacting matter undergoes a smooth  crossover transition from the hadronic phase to the quark-gluon plasma, where the spontaneously broken chiral symmetry is restored~\cite{Aoki:2006we,Borsanyi:2010bp}. The order of the transition in the limit of massless u and d  quarks is a subtle issue, and still under debate~\cite{Bonati:2014kpa}. Here we assume that the transition in this limit is second order, belonging to the $O(4)$ universality class.   Owing to the sign problem, the nature of the transition  at higher  net-baryon  densities is not settled by first principle LQCD studies. However, in  effective models of QCD, it is found  that, at sufficiently large $\mu_B$,  the systems can exhibit a first order  chiral phase  transition. The endpoint of this conjectured transition line  in the  $(T,\mu_B)$- plane, is the chiral {\sl critical point} (CP) \cite{Asakawa:1989bq,Halasz:1998qr}.  At the CP, the system exhibits a 2$^{nd}$ order phase transition, which belongs to the $Z(2)$ universality class \cite{Schaefer:2006ds}.

Due to the restriction  of present  LQCD calculations to small net-baryon densities, effective models  that belong to the same universality class as QCD, e.g., the Polyakov-loop extended,  Nambu-Jona-Lasinio (PNJL) \cite{Fukushima:2003fw,Ratti:2005jh,Sasaki:2006ww} and Quark-Meson (PQM)  models \cite{Schaefer:2007pw,Schaefer:2009ui,Herbst:2010rf,Skokov:2010wb,Skokov:2010sf,Skokov:2010uh}, have been  employed to study the chiral  phase transition
for a broad range of thermal parameters.

One of the strategic goals of current experimental and theoretical studies of chiral symmetry restoration in QCD is to unravel the phase diagram of strongly interacting matter and to clarify whether a chiral CP exists. A dedicated research program  at RHIC, the beam energy scan, has been established to explore these issues in collisions of heavy ions at relativistic energies \cite{Aggarwal:2010cw}. By varying the beam energy  at RHIC, the properties of strongly interacting matter in a broad range of net-baryon densities, correponding to a wide range in baryon chemical potential, ${20\;\mathrm{MeV} <\mu_B<500\; \mathrm{MeV}}$ \cite{BraunMunzinger:2003zd,Cleymans:2005xv}, can be studied.
To study the phase structure, the fluctuations of conserved  charges have been proposed as probes~\cite{Stephanov:1998dy,Stephanov:1999zu,Asakawa:2000wh,Jeon:2000wg,Friman:2011pf}.
These  are experimentally accessible and reflect the criticality of the chiral transition. 

First data on net-proton-number fluctuations, which are used as a proxy for fluctuations of the  net-baryon number,  have been obtained in heavy-ion collisions by the STAR Collaboration at  RHIC energies \cite{Adamczyk:2013dal,Luo:2015ewa,Luo:2015doi}. The STAR  data  on the  variance, skewness and kurtosis of net proton number,  are intriguing and have stimulated a lively discussion on their physics origin and interpretation \cite{Proceedings:2016arp}.

In the following,
we focus on  the properties and systematics  of the  cumulants of net-baryon-number  fluctuations near the chiral phase boundary. Our studies are performed
in the Polykov-loop extended  Quark-Meson  model which includes  coupling of quarks to  temporal gauge fields. To account for the  $O(4)$ and $Z(2)$ critical fluctuations
near the phase boundary, we employ the Functional Renormalization Group \cite{Wetterich:1992yh,Morris:1993qb,Berges:2000ew}.
We present results for ratios of cumulants obtained on the phase boundary and  on a freezeout line determined by fitting the skewness ratio, following \cite{Bazavov:2012vg, Ournewpaper}. These are confronted with the corresponding experimental values of the STAR collaboration.

\section{The Polyakov--quark-meson model \label{sec:PQM}}
The PQM model is a low energy effective approximation  to QCD  formulated in terms of  the light quark  $q=(u,d)$ as well as scalar and the pseudoscalar meson $\phi=(\sigma,\vec{\pi})$ fields. The quarks are coupled to  the  background Euclidean gluon field $A_{\mu}$, with vanishing  spatial components,  which is linked  to  the Polyakov loop
\begin{equation}
\Phi =            \frac{1}{N_c}\left\langle \mathrm{Tr}_c \mathcal{P} \exp\left( i\int_0^{\beta} d\tau A_0 \right) \right\rangle .
\end{equation}
The resulting Lagrangian of the model reads
\begin{align}\label{lagrangian}
\begin{split}
\mathcal{L} &= \bar{q}\left( i\gamma^{\mu} D_{\mu} -g(\sigma +i\gamma_5 \vec{\tau}\vec{\pi})\right)q+\frac12 \left(\partial_{\mu}\sigma\right)^2+\frac12 \left(\partial_{\mu}\vec{\pi}\right)^2
\\
&-U_m(\sigma,\vec{\pi})-\mathcal{U}(\Phi,\bar{\Phi};T)-\frac12 m_{\omega}^2\omega^2,
\end{split}
\end{align}
where $D_{\mu}=\partial_{\mu}-iA_{\mu}$. The parameters of the mesonic potential
\begin{equation}
U_m(\sigma,\vec{\pi}) = \frac{\lambda}{4}(\sigma^2+ \vec{\pi}^2)^2 +\frac{m^2}{2}(\sigma^2+ \vec{\pi}^2) - H\sigma,
\end{equation}
are tuned to vacuum properties of the $\sigma$ and $\vec{\pi}$ mesons.
We use the Polyakov-loop potential $\mathcal{U}(\Phi,\bar{\Phi};T)$ determined in \cite{Lo:2013hla}.

We compute the thermodynamics of this model,
including fluctuations of the scalar and pseudoscalar meson fields within the framework of the FRG method. The Polyakov loop is treated on the mean-field level. Its value its tuned such that a stationary point of the thermodynamic potential is reached at the end of the RG calculation.


In the FRG framework, the effective average action $\Gamma_k$, which  interpolates between the classical and the full quantum action, is obtained by solving the renomalization group flow equation~\cite{Wetterich:1992yh}
\begin{equation}\label{eq:FRG-flow}
\partial_k \Gamma_k[\phi]=\frac12 \mathrm{Tr}\left[\left( \Gamma_k^{(2)}[\phi]+R_k \right)^{-1} \partial_k R_k\right],
\end{equation}
where $\phi$ denotes the quantum fields considered, $\mathrm{Tr}$ is a trace over the fields, over momentum  and over all internal indices. The regulator function $R_k$ suppresses fluctuations at momenta below $k$. Thus, effects of fluctuations of quantum fields are included shell by  shell in  momentum space, starting from a UV cutoff scale $\Lambda$. We employ the  optimized regulator introduced by Litim \cite{Litim:2001up}. Details of the calculation can be found in~\cite{Ournewpaper}.

\section{Net-baryon-number cumulants and  the phase boundary}

The chiral Lagrangian introduced above shares the chiral critical properties with QCD. In particular,  at moderate values of the chemical potential, the PQM model exhibits a chiral transition belonging to the $O(4)$ universality class. For larger values of $\mu$, it reveals a $Z(2)$ critical endpoint, followed  by a first order  phase transition \cite{Schaefer:2006ds}. Consequently, the PQM model embodies the generic  phase structure expected for QCD,  with the universal $O(4)$ and $Z(2)$ criticality  encoded in the scaling functions. Furthermore, due to the coupling of the quarks to the background gluon fields, the PQM model incorporates  "statistical confinement", i.e., the suppression  of quark and diquark degrees of freedom in the low temperature,  chirally broken phase~\cite{Fukushima:2003fw}. Consequently,  by studying fluctuations of conserved charges in the PQM model, one  can explore the influence of chiral symmetry restoration and of "statistical confinement"  on the cumulants  in  different sections   of  the chiral phase boundary. The baryon- and quark-number cumulants of order $n$, $\chi_B^n$ and $\chi_q^n$, and the baryon-number cumulant ratios, $\chi_B^{n,m}$, are defined as
\begin{equation}\label{eq:chidef}
\chi_B^n =\frac{\chi_q^n}{3^n} = \frac{1}{3^n T^{4-n}}\frac{\partial^n \Omega(T,\mu_q)}{\partial \mu_q^n},\quad\quad
\chi_B^{n,m}=\frac{\chi_B^{n}(T,\mu_B)}{\chi_B^{m}(T,\mu_B)}.
\end{equation}
In the following we focus on ratios of net-baryon-number susceptibilities that can be related to experimentally measurable quantities:
\begin{equation}
\chi_B^{1,2}(T,\mu_B)=\nicefrac{M}{\sigma^2},~~
\chi_B^{3,1}(T,\mu_B)=\nicefrac{S_B\sigma^3}{M},~~
\chi_B^{4,2}(T,\mu_B)=\kappa\sigma^2
\label{def}
\end{equation}
where $M$ is the mean, $\sigma$ the variance, $S_B$ the skewness and $\kappa$ the kurtosis of the net-baryon-number distribution.

\begin{figure}[tb]
	\includegraphics[width=0.49\linewidth]{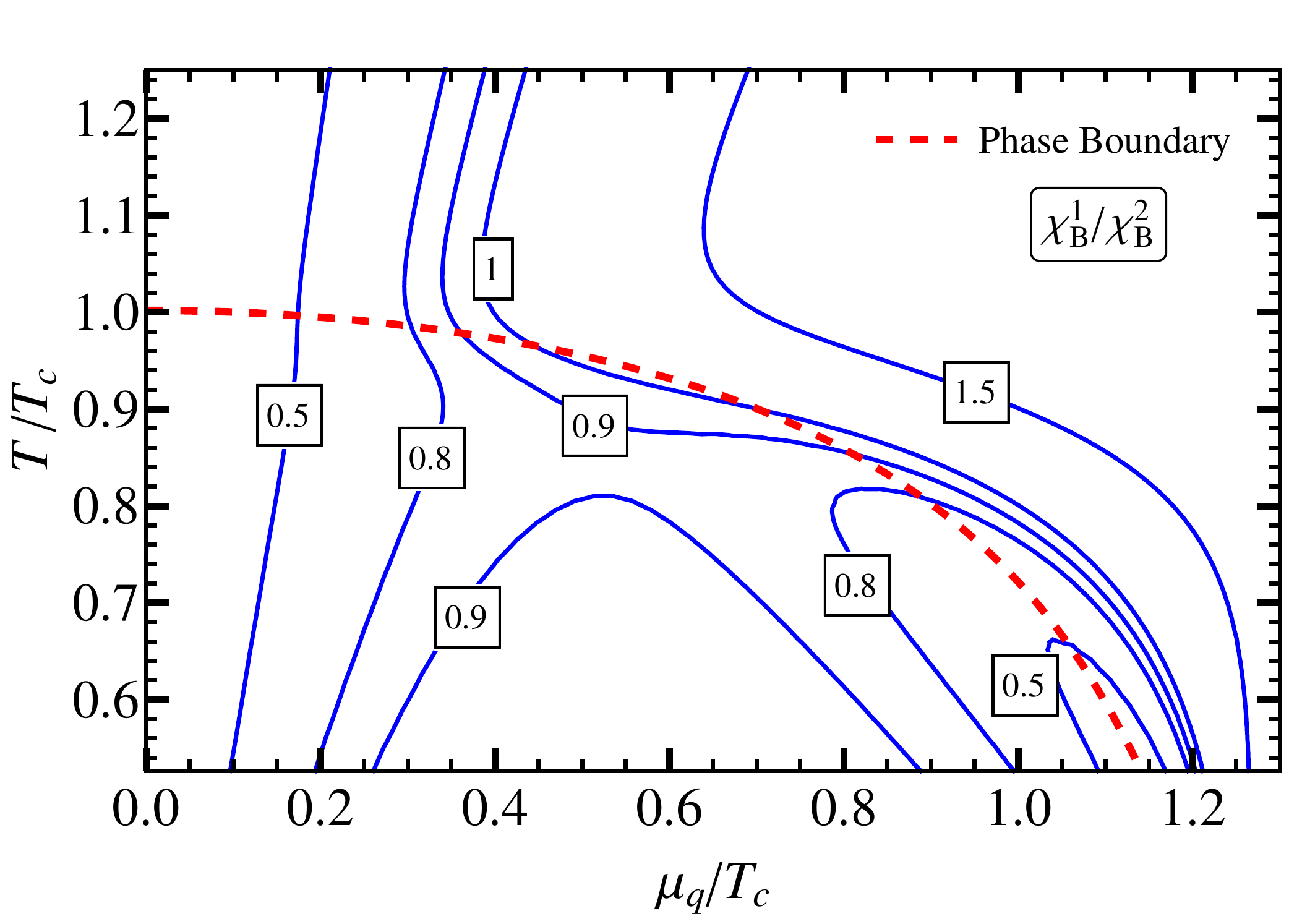}
	\includegraphics[width=0.49\linewidth]{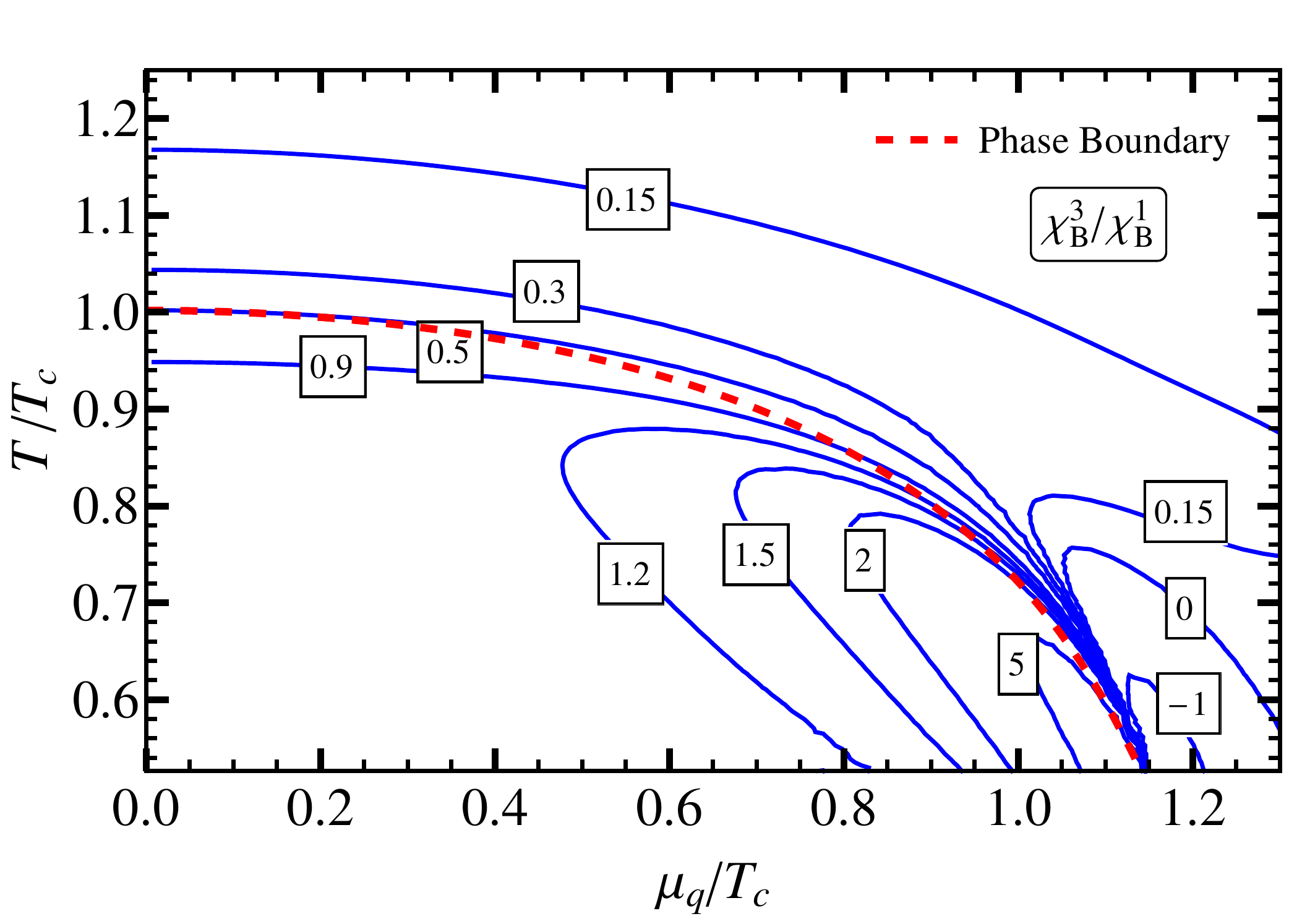}
	\caption{Contour plots of the ratios  $\chi_B^1/\chi_B^2$  and $\chi_B^3/\chi_B^1$   in the $(T,\mu_q)$-plane, computed in the PQM model. The broken lines indicate the location of the chiral crossover phase boundary. \label{fig:c1c2}}
\end{figure}
\begin{figure}[tb]
	\includegraphics[width=0.49\linewidth]{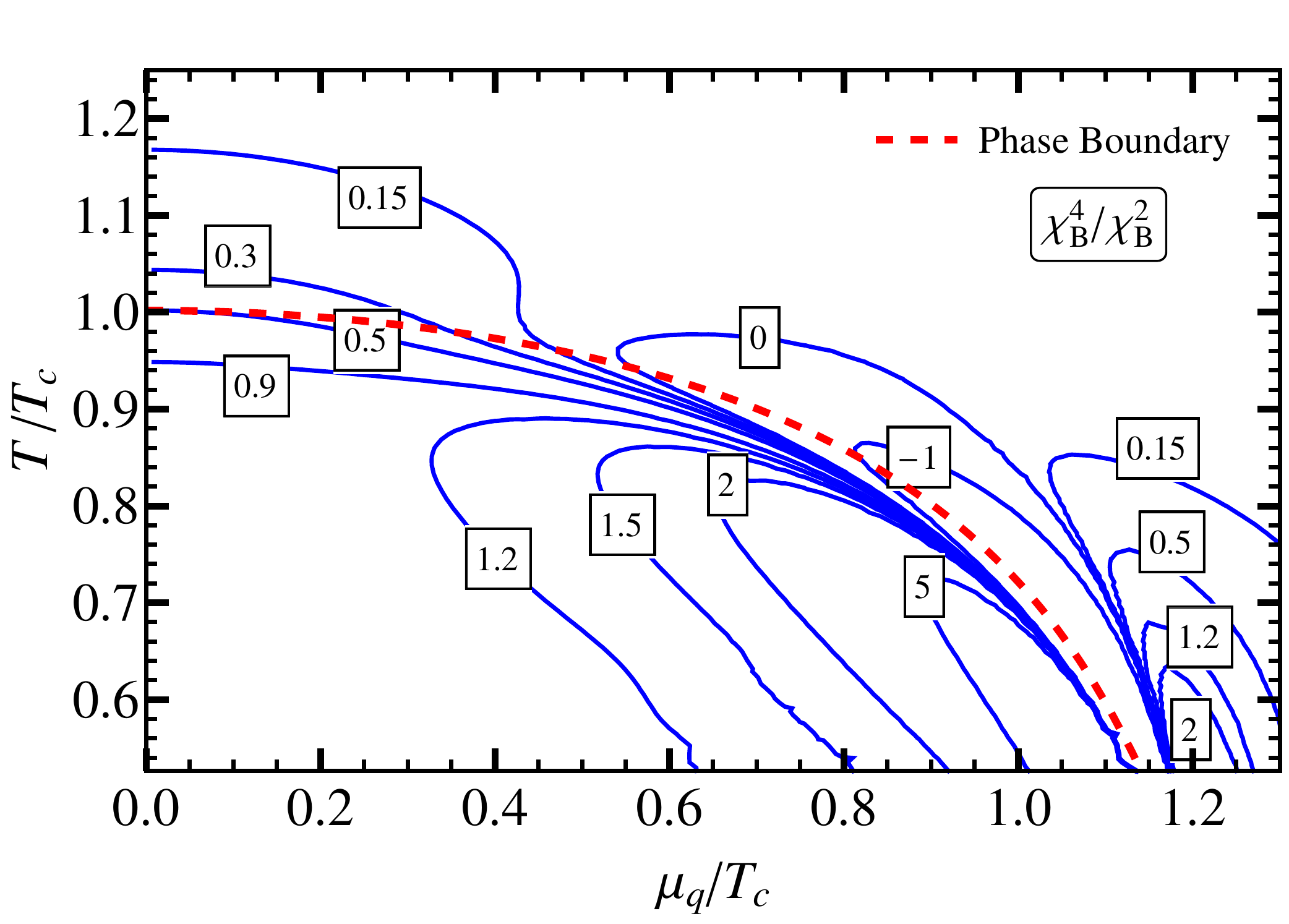}
   \includegraphics[width=0.48\linewidth]{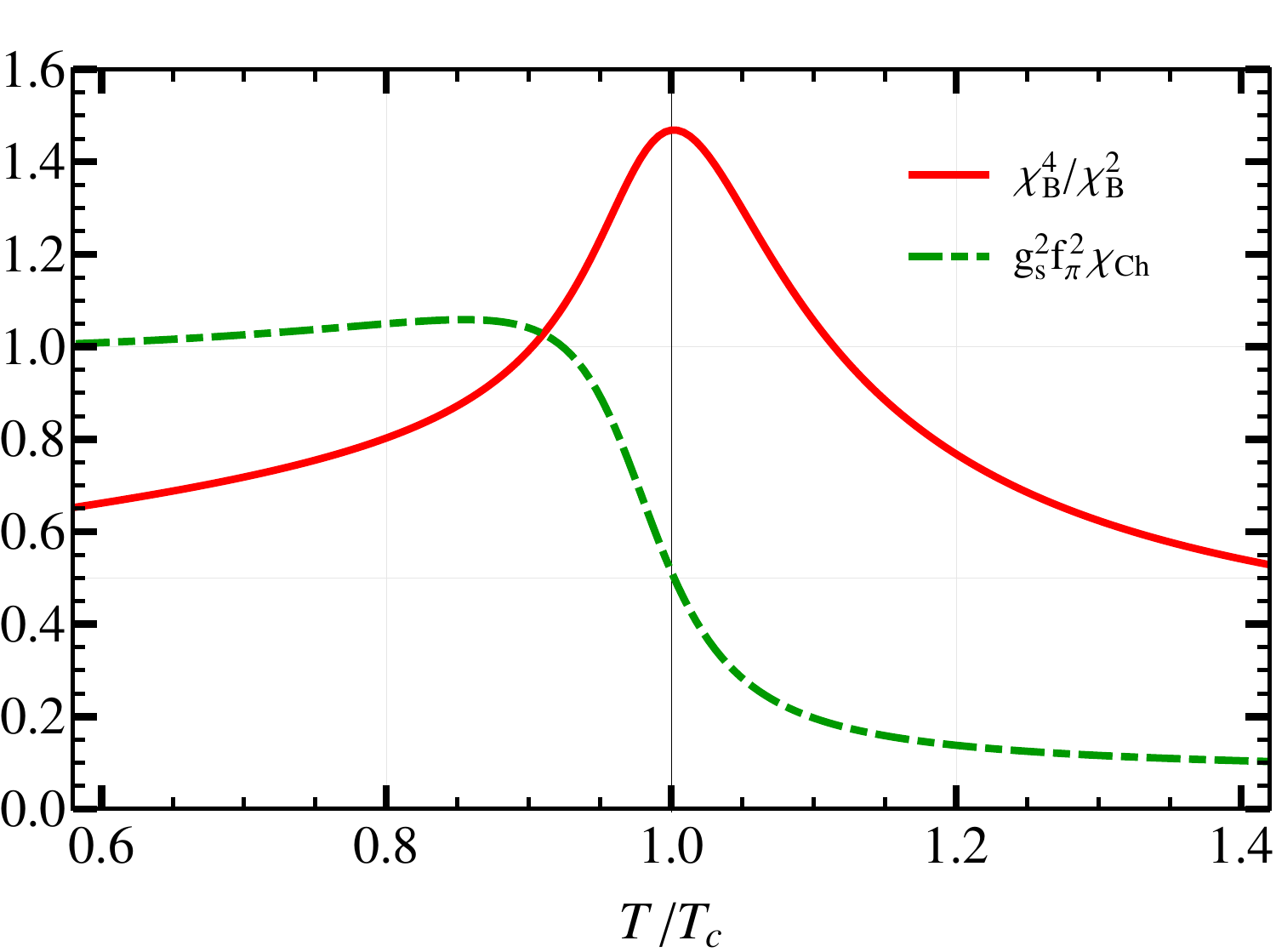}
	\caption{Left: contour plot of the kurtosis ratio, $\chi_B^4/\chi_B^2$,
		in the $(T,\mu_q)$ plane. Right: temperature dependence of ratios of net-baryon-number   cumulants, ${\chi_B^3/\chi_B^1=\chi_B^4/\chi_B^2}$ in the PQM model at $\mu_q=0$.
		\label{fig:c4c2}}
\end{figure}

At vanishing chemical potential, all odd susceptibilities of the net baryon number vanish, owing to the baryon-antibaryon symmetry. In addition, in the $O(4)$ universality class, the second and fourth order cumulants remain finite at the phase transition temperature  at $\mu_q=0$ in the chiral limit, implying that only sixth and higher order susceptibilities  diverge. Thus, for physical quark masses, only higher order cumulants,   $\chi_B^n$ with $n>4$, can exhibit $O(4)$ criticality at $\mu_q=0$~\cite{Friman:2011pf}. A further consequence of the
baryon-antibaryon symmetry is the equality of the ratios 
\begin{equation}
\chi_B^{2m-1,2n-1}=\chi_B^{2m,2n}
\label{eq:odd-even}
\end{equation}
for any integer $m$ and $n\,\geq 1$ at $\mu_q=0$. For $\chi_B^{3,1}$ and $\chi_B^{4,2}$, the equality at small $\mu_q$ can also be seen by comparing the right panel of Fig.~\ref{fig:c1c2} to the left panel of Fig.~\ref{fig:c4c2}.


At finite net-baryon density the singularity at the $O(4)$ critical line is stronger than at $\mu_q=0$. Thus, in this case the third-order cumulant and all higher-order ones diverge at the $O(4)$ line. The second order cumulant $\chi_B^2$  remains finite, and  diverges at the tricritical point and for non-zero quark masses, at the CP.

In  Fig.~\ref{fig:c1c2} we show contour plots of the ratios $\chi_B^{1,2}$ and $\chi_B^{3,1}$  in the $(T,\mu_q)$ plane. As noted above, all odd cumulants vanish at $\mu_q=0$. Consequently, $\chi_B^{1,2} |_{\mu_q=0}=0$ for any $T$, while the ratio
$\chi_B^{3,1} |_{\mu_q=0}$ is non-vanishing. As  indicated in Fig. \ref{fig:c1c2}, the ratio  $\chi_B^{3,1} $ decreases with temperature, and depends weakly on the chemical potential. Thus, the ratio $\chi_B^{3,1}$ can be used as a measure of the temperature. 

In Fig. \ref{fig:c4c2} we show contour plots of the ratios $\chi_B^{4,2}$ and results on the temperature dependence of the ratio $\chi_B^{3,1}=\chi_B^{4,2}$  of net-baryon-number susceptibilities at $\mu_q=0$,    together with the variance  of the chiral condensate,  $\chi_{ch}$.
The location of the maximum of the  chiral susceptibility, $\chi_{ch}$, defines  the pseudo-critical temperature,  $T_c$.

At small $\mu_q/T$, the properties of the first four susceptibilities,  $\chi_B^n$ with $n=1,..,4$,  and consequently their ratios, near the chiral crossover are dominantly affected by the  coupling of the quarks to the Polyakov loop,  and the resulting statistical confinement. The critical chiral dynamics, i.e.  the  $O(4)$ and $Z(2)$ criticality at the chiral crossover transition and at the CP, respectively, unfolds at larger $\mu/T$.  Near the CP, there is a strong variation of the cumulants  with $T$ and $\mu_q$,  which   increases  with the order of cumulants. 

\section{Net-baryon  cumulant ratios and freeze-out in heavy ion collisions\label{sec:FO}}

In heavy-ion collisions, the thermal fireball  formed  in the   quark-gluon plasma phase
undergoes expansion and passes through the QCD phase boundary  at some point   $(\mu_q,T)$, which depends on  the collision energy,  $\sqrt s$. Analysis of ratios of particle multiplicities indicate that at high beam energies (small values of $\mu_q/T$), the freeze out occurs at or just below the phase boundary.  Thus, the beam energy dependence of net-baryon-number  susceptibilities  can provide insight into the structure of the QCD phase diagram and information on the existence and location  of the  CP. Consequently, it is of phenomenological interest to compute the    properties of  fluctuations of conserved  charges along the chiral phase boundary. Since there,  the  critical structure and the relations between different susceptibilities are by and large governed by the universal scaling functions, the   generic behavior of  ratios of net-baryon-number susceptibilities  can be explored  also in  model calculations.

\begin{figure}[tb]
	\includegraphics[width=0.49\linewidth]{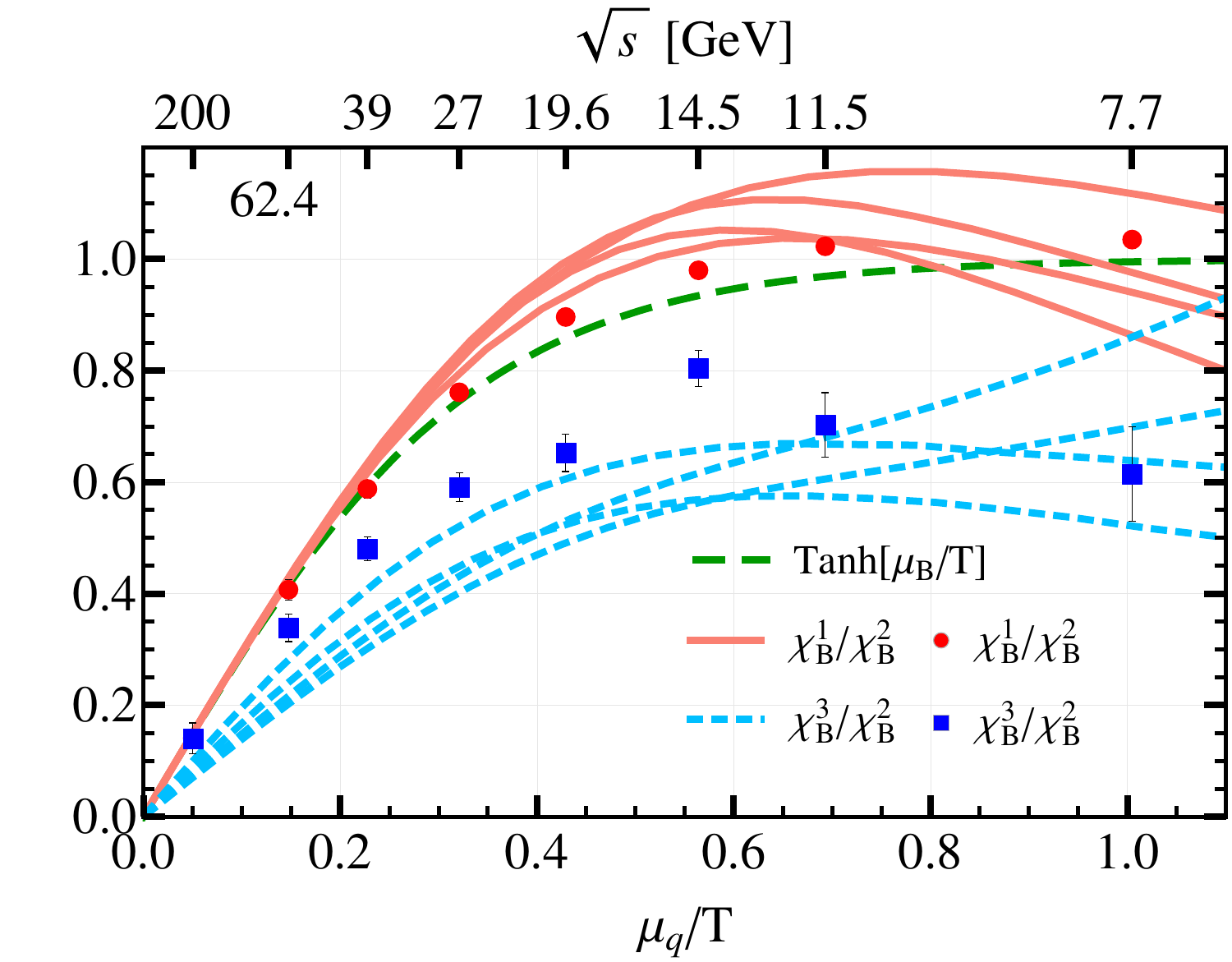}
	\includegraphics[width=0.49\linewidth]{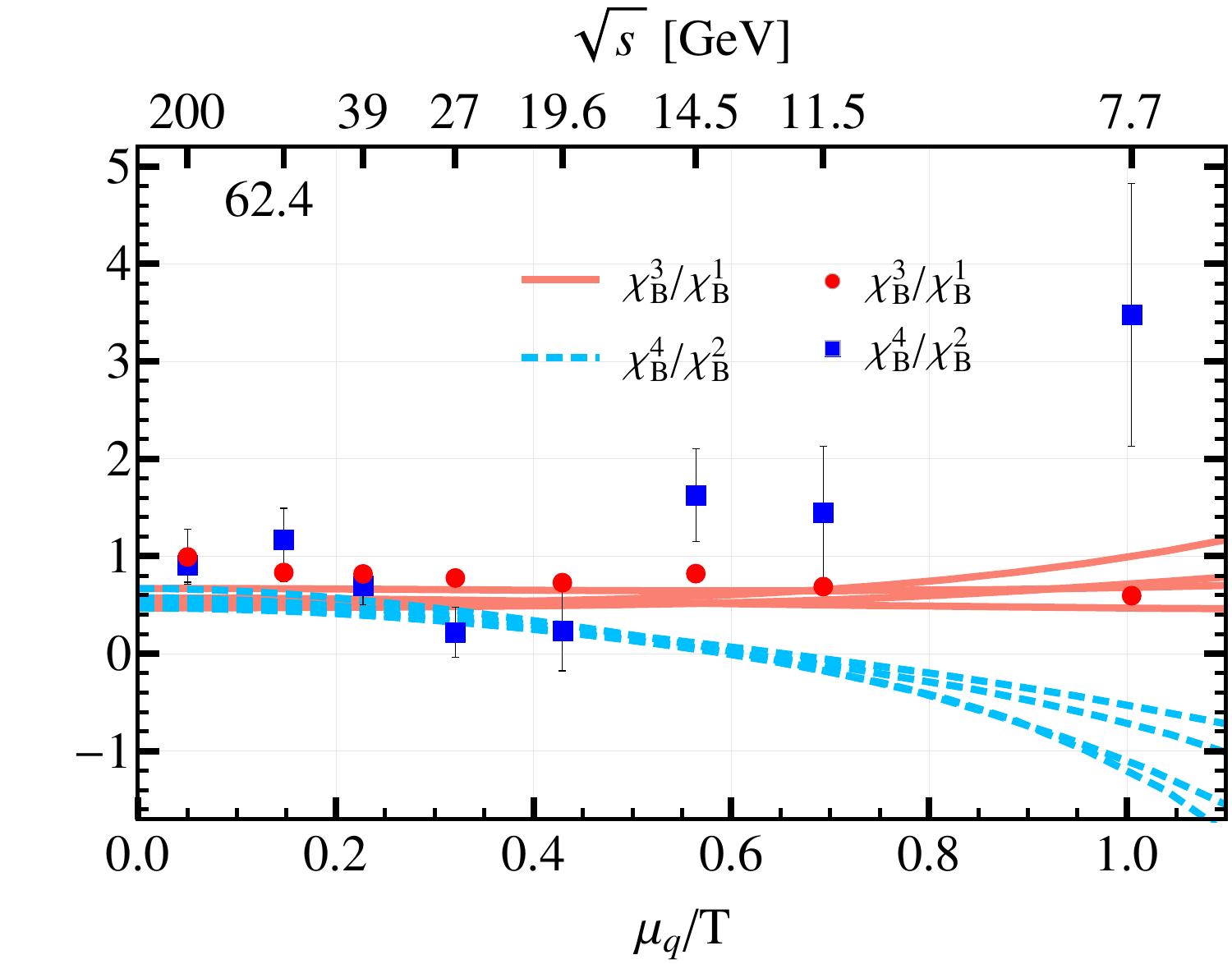}
	\caption{
		Ratios of cumulants of net-baryon-number fluctuations  in the PQM model, computed along the chiral phase boundary,  for four sets of model parameters~\cite{Ournewpaper}. Also shown are the preliminary STAR data~\cite{Luo:2015ewa,Luo:2015doi}, assuming the relation between the ratio $(\mu_q/T)$  and the collision energy obtained by analysing the chemical freezout conditions \cite{Andronic:2005yp,Andronic:2008gu,Cleymans:2005xv}. The green dashed line in the left figure shows the baseline result for $\chi_B^{1,2}$, $\tanh(3\mu_q/T)$.
		\label{fig:five}}
\end{figure}

\begin{figure}[tb]
	\includegraphics[width=0.49\linewidth]{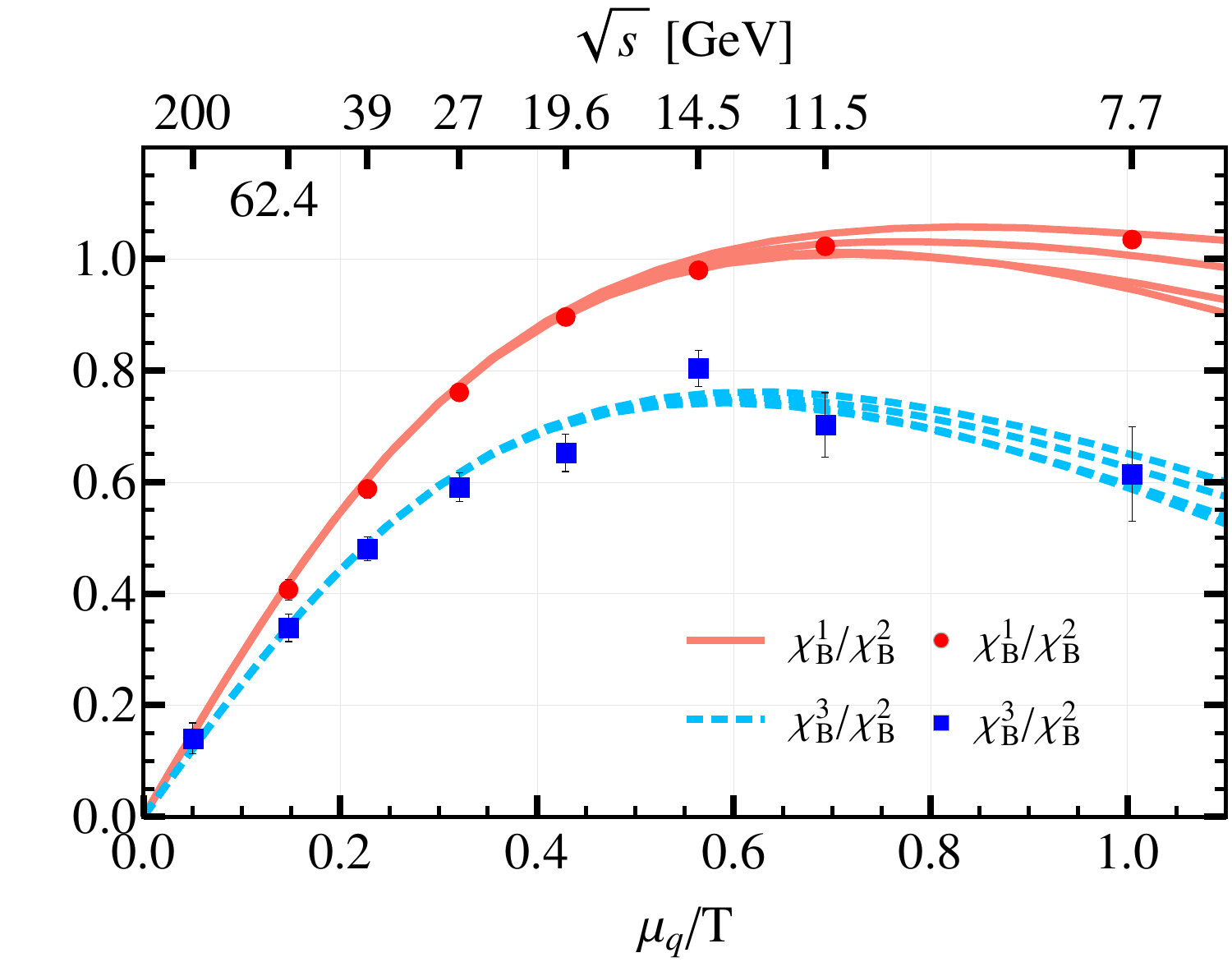}
	\includegraphics[width=0.49\linewidth]{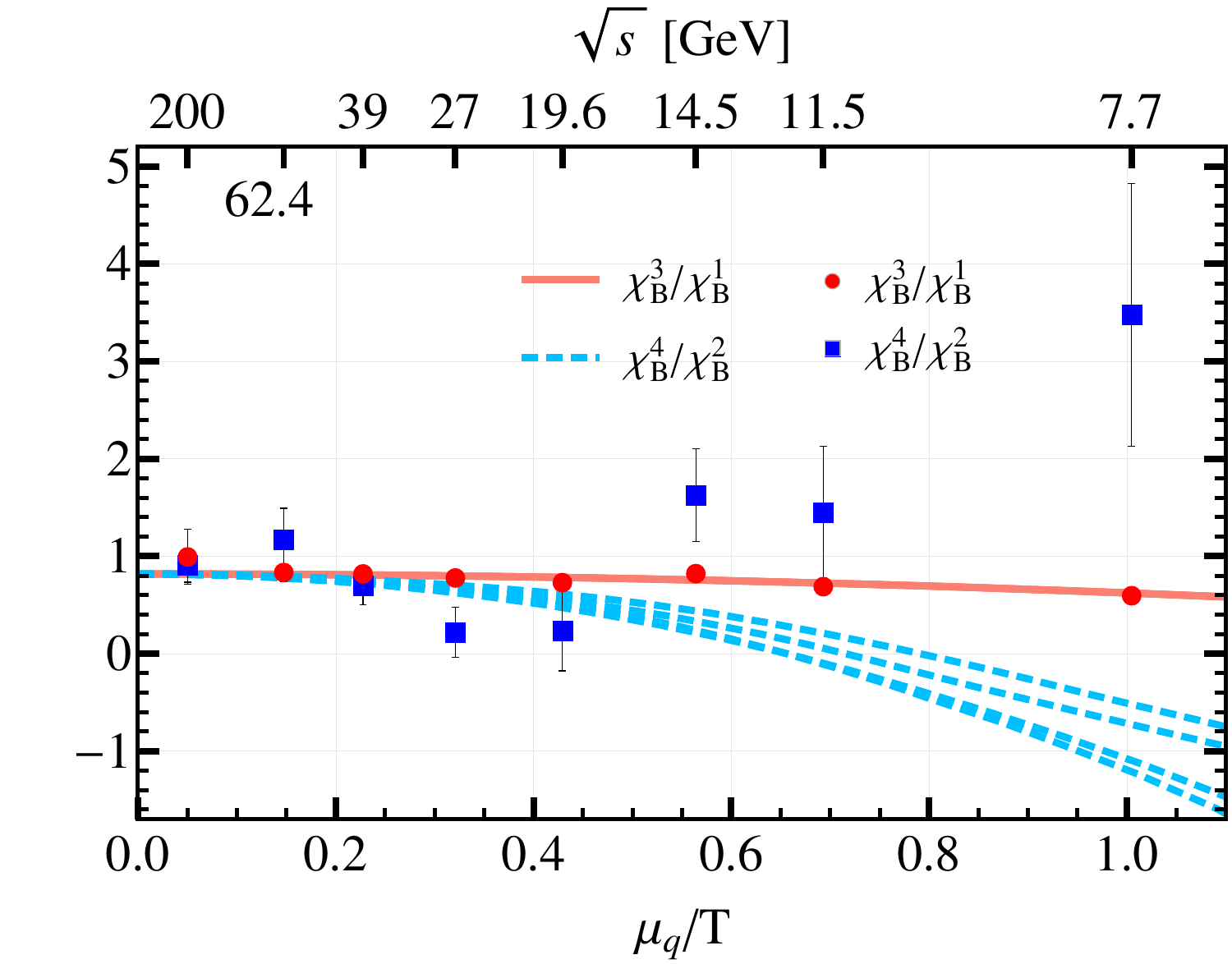}
	\caption{
		Ratios of cumulants of net-baryon-number fluctuations  in the PQM model  along the freezeout line, obtained by fitting $\chi_B^3/\chi_B^1$ to the STAR data.  The four sets of model parameters used and the preliminary STAR data shown, are the same as in Fig. \ref{fig:five}. 
		\label{fig:six}}
\end{figure}

A comparison of results obtained in the PQM model with data, requires a correspondence between the collision energy $\sqrt s$ and the thermal parameters $(\mu_q, T)$. Here we employ the phenomenological relation,  obtained by analysing the freeze-out conditions in terms of the hadron-resonance-gas model (HRG) \cite{BraunMunzinger:2003zd,Cleymans:2005xv}. We then  use the  resulting dependence of $\mu_B$ and $T$ on $\sqrt s$ to assign a value for the ratio $\mu_q/T$ to each of the STAR beam energies. We note that, for $\mu_q/T<1$,  the phenomenological   freeze-out line coincides within errors with the crossover  phase boundary obtained in   lattice QCD  \cite{Borsanyi:2010bp,Andronic:2008gu}. This  motivates a comparison of model results  on net-baryon-number  fluctuations near  the phase boundary with data. A similar analysis was first done using LQCD results in Ref.~\cite{Bazavov:2012vg}.

In Fig.~\ref{fig:five}, we show the STAR data on net-proton-number  susceptibility ratios and the corresponding PQM model results on net-baryon-number fluctuations computed along the phase boundary. The model results for the ratios $\chi_B^{1,2}$, $\chi_B^{3,1}$ and $\chi_B^{3,2}$ are in qualitative agreement with the data in the whole energy range. For the kurtosis ratio, $\chi_B^{4,2}$, this is the case  also  up to the SPS energy, i.e., for $\sqrt s \geq 20$ GeV. However, for $\mu_q/T>0.5$,  the data on the kurtosis ratio exhibits a qualitatively different dependence on $\mu_q/T$  than  expected for the critical behavior of $\chi_B^{4,2}$, as the CEP is approached along the phase boundary.

In the comparison of model predictions with data in Fig.  \ref{fig:five}, we assume,  that the freezout of the net-baryon-number  fluctuations tracks the chiral phase boundary. Clearly, this assumption provides a qualitative understanding of the data. In order to obtain a more quantitative description, we follow
Refs.  \cite{Bazavov:2012vg,Almasi:2016hhx,Fu:2015amv}, and  determine the  freezout conditions by fitting the data on the $\chi_B^{3,1}$ ratio, using the $\sqrt s$-dependence  of $\mu_q/T$ obtained from  the fit of the  HRG model to particle multiplicities \cite{Andronic:2005yp,Andronic:2008gu,Cleymans:2005xv}.

In  Fig. \ref{fig:six} we show the fluctuation ratios along the freezeout line that is fixed through the skewness  data. The model results are obtained for four sets of initial conditions introduced in~\cite{Ournewpaper}. Fig. \ref{fig:six} clearly shows that, along the freezeout line, the spread of all fluctuations ratios considered  for the various parameter sets is much weaker than that observed in Fig. \ref{fig:five} along the phase boundary.  This indicates that moderate changes of the sigma mass and modifications of the form of the Polyakov loop potential may lead to a shift in the temperature scale but essentially no change of the relative structure of the cumulant ratios.

The results presented in Fig. \ref{fig:six} clearly show that the model provides a very good description of the data on $\chi_B^{1,2}$ and   $\chi_B^{3,2}$. Also the  kurtosis data,  obtained at higher  collision energies,  are consistent with model results.  However,  at $\sqrt s<  20$ GeV the latter again exhibits a different  trend, with the data increasing rapidly at lower energies, while the model result keeps decreasing.


The comparison of model results on ratios of net-baryon-number susceptibilities with the STAR data in Figs. \ref{fig:five} and \ref{fig:six} shows that the data, with the exception of kurtosis at low energies, follow  general trends expected due to critical chiral dynamics and general considerations. We note that the ratios of net-baryon-number susceptibilities  near the phase boundary involving net-baryon-number cumulants $\chi_B^n$ with $n\geq 3$  are  controlled mainly by the  scaling functions in the $O(4)$ and $Z(2)$ universality classes, respectively. This observation indicates, that by measuring fluctuations of conserved charges in heavy-ion collisions,  we are indeed probing the QCD phase boundary, and thus accumulating evidence for chiral symmetry restoration.

However, as discussed above, there are several uncertainties and assumptions which must be thoroughly understood before the QCD phase boundary can be pinned down with confidence. Possible contributions to fluctuation observables from effects not related to critical phenomena, like e.g. baryon-number conservation~\cite{Bzdak:2012an}  and volume fluctuations~\cite{Skokov:2012ds,Braun-Munzinger:2016yjz} are being explored. We mention in particular the  rather  strong sensitivity  of higher order net-proton-number cumulants on the transverse momentum range imposed in the analysis of the STAR data.  Nevertheless, it is intriguing  that the dynamics of this model provides a good description of the STAR data (except for $\chi_B^4$ at the lowest energies), without all these effects of non-critical origin. It remains an important task to assess the effect of theses additional sources of fluctuations in the whole energy range probed by the experiments.

\section*{Acknowledgments}
We acknowledge stimulating discussions with Peter Braun-Munzinger, Frithjof Karsch and  Nu Xu. The work of B.F. and K.R. was partly supported
by the Extreme Matter Institute EMMI.
K. R. also  acknowledges  partial  supports of the Polish National Science Center (NCN) under Maestro grant DEC-2013/10/A/ST2/00106.
G. A. acknowledges the support of  the Hessian LOEWE initiative
through the Helmholtz International Center for FAIR (HIC for FAIR).

\bibliographystyle{apsrev4-1}
\bibliography{refs}
\end{document}